\documentclass[12pt,preprint]{aastex}

\newcommand{\sax}{{\it BeppoSAX} }

\newcommand{\integral}{{\it INTEGRAL} }
\newcommand{\chandra}{{\it Chandra} }

 \def\source{SAX~J1810.8--2609}
\def\be{\begin{equation}} 
\def\ee{\end{equation}} 
 
\begin{document}

\title{Renewed activity from the X-ray transient SAXJ~1810.8-2609
      with \integral\/ \altaffilmark{1}}

\altaffiltext{1}{INTEGRAL is an ESA project with instruments 
 and science data center funded by ESA member states 
(especially the PI countries: Denmark, France, Germany, Italy, Switzerland, Spain),
 Czech Republic and Poland, and with the participation of Russia and the USA.} 

\author{M. Fiocchi\altaffilmark{1}, L. Natalucci\altaffilmark{1}, J. Chenevez\altaffilmark{2}, A. Bazzano\altaffilmark{1}, A. Tarana\altaffilmark{1}, P. Ubertini\altaffilmark{1}, S. Brandt\altaffilmark{2}, V. Beckmann\altaffilmark{3}, M. Federici\altaffilmark{1}, R. Galis\altaffilmark{3,4}, R. Hudec\altaffilmark{3,4} }
\altaffiltext{1}{Istituto di Astrofisica Spaziale e Fisica Cosmica di Roma (INAF). Via Fosso del Cavaliere 100, Roma, I-00133, Italy} 
\altaffiltext{2}{National Space Institute, Technical University of Denmark, Juliane Maries Vej 30, 2100 Copenhagen, Denmark
}
\altaffiltext{3}{INTEGRAL Science Data Centre, Chemin d'Ecogia 16, CH-1290 Versoix, Switzerland
}
\altaffiltext{4}{Astronomical Institute, Academy of Sciences of the Czech Republic, CZ-251
 65 Ondrejov, Czech Republic}

\begin{abstract}

We report on the results of 
\integral\/ observations
of the neutron star 
low mass X-ray binary SAX~J1810.8-2609 during its latest active phase
in August 2007.
The current outburst is the first one since 1998 and the derived luminosity 
is 
$1.1-2.6~\times~10^{36} \rm \, erg \, s^{-1}$
in the 20--100 keV energy range.
This low outburst luminosity and the long-term time-average 
accretion rate of $\sim~5\times10^{-12} {\rm M}_{\sun}$ yr$^{-1}$
suggest that \source\/ is a faint soft X-ray transient.
During the flux increase, spectra are consistent with a thermal
Comptonization model with a temperature plasma of 
$kT_e \sim$ 23-30 keV 
and an optical depth of 
$\tau \sim$ 1.2-1.5, independent from luminosity of the system.
This is a typical low hard spectral state for which the X-ray emission is attributed to
the upscattering of soft seed photons by a hot, optically thin electron plasma. 
During the decay, spectra have a different shape, the high energy tail being 
compatible with a single power law.  
This confirm similar behavior observed by \sax\/ during the previous outburst,
with absence of visible cutoff in the hard X-ray spectrum.
\\
{\it INTEGRAL}/JEM-X instrument observed four X-ray bursts  
 in Fall 2007.
The first one has the highest peak flux ($\approx3.5~Crab$ 
in 3--25 keV)
giving an upper limit to 
the distance of the source of about 5.7~kpc, for a  
$L_{\rm Edd}\approx\,3.8\times 10^{38}$ erg s$^{-1}$.
The observed recurrence time of $\sim$1.2 days and
the ratio of the total energy emitted in the persistent flux to that 
emitted in the bursts ($\alpha\sim$73) allow us to conclude that the burst fuel 
was composed by mixed hydrogen and helium with X$\geq$0.4. 

\end{abstract}


\section{Introduction}
The transient X-ray source SAX~J1810.8-2609 was discovered on 1998 March 10 (Ubertini et al. 1998) with the Wide Field Cameras (2-28 keV)  
on-board the {\it BeppoSAX} satellite. 
During the performed Galactic Bulge monitoring, a strong type I X-ray burst was detected identifying 
this compact object as a neutron star in a low-mass X-ray binary system. Assuming standard burst parameters 
and attributing the photospheric radius expansion to near-Eddington luminosity, 
the distance was estimated be $\sim~5$ kpc (Natalucci et al. 2000).
The wide-band spectral data (0.1--200 keV) obtained later with the NFI/\sax\/ showed a hard X-ray spectrum described by 
a power law with photon spectral index $\Gamma=1.96\pm0.04$ 
and a soft component which was compatible with blackbody radiation 
of temperature kT$\sim$0.5 keV (Natalucci et al. 2000).

During a 
{\it ROSAT} follow-up observation on 1998 March 24 an 
X-ray source (named RX J1810.7--2609)
was detected at a position consistent with the WFC error box (Greiner et al. 1998). 
Optical-to-infrared follow-up observations of the 10$''$ radius ROSAT HRI X-ray error box
revealed one variable object 
(R=$19.5\pm0.5 \rm \, mag$ on March 13, R$>21.5 \rm \, mag$ 
on 1998 August 27) which was proposed as the optical/IR counterpart of RX~J1810.7--2609 and SAX~J1810.8--2609 (Greiner et al. 1998).

Using \chandra\/ instruments, Jonker et al. (2004) detected the neutron star 
system in quiescence at an 
unabsorbed luminosity of 
$\sim 10^{32} \rm \, erg \, s^{-1}$
(assuming a distance of 4.9 kpc). 
The quiescent spectrum is well-fitted with an absorbed power law with a photon index 
$\Gamma = 3.3\pm0.5$ and the Galactic absorption value ($N_{H,gal} =
3.3 \times 10^{21} \rm \, cm^{-2}$)
consistent with the value derived in outburst.

Since 1998 this burster remained in a quiescent state. 
Only in August 2007 
{\it Swift} 
observed a new phase of activity 
(Parson et al. 2007). The {\it Swift}/UVOT
instrument detected 
a weak source in the 
white-band filter at the position of 
SAX~J1810.8--2609 and did not detect it in any other single filter
(Scady et al. 2007). The source was observed on a daily basis with
{\it Swift} 
using $\sim 1ksec$ exposure,
starting August 6, 2007. In all observations the 
{\it Swift}/XRT $0.3-10 \rm \, keV$  
spectrum was well fitted using 
an absorbed power law model with a hydrogen column density of 
$N_{H} \sim~5\times10^{21} \rm \, cm^{-2}$ and a spectral index of $\Gamma \sim~2$ (Degenaar et al. 2007). 
After few months, the source went back to quiescent state, indeed on
November 3rd and 5th, 
{\it Swift}/XRT did not detect it during two individual $\sim~1.6 \rm
\, ksec$ and $\sim~1.9 \rm \, ksec$ observations. 

The X-ray spectra of LMXBs are usually fit with a complex model:
at low energies a blackbody component that approximates the spectrum of
an optically thick, geometrically thin accretion disk
and/or the neutron star surface, and at higher energies a
Comptonization component due to repeated inverse Compton scattering of the soft seed
photons by hot electrons plasma with a thermal distribution of velocities.
\sax\/ and \integral\/ results showed that the hard component can extend up to 200 keV 
without any appreciable break (Di Salvo et al. 2000, 2001, Fiocchi et al. 2006, 
Piraino et al 1999, Iaria et al. 2001, Tarana et al. 2006). 
In this paper, we study the spectral behavior of the X-ray transient burster \source\/,
showing this behavior during the decay of the outburst. 
Finally, we report on four X-ray bursts observed by INTEGRAL/JEM-X instrument in Fall 2007.

\section{OBSERVATIONS AND DATA ANALYSIS}
The \integral\/ (Winkler et al. 2003) observations
are divided into uninterrupted 2000~s intervals, the so-called science windows (SCWs).
Spectra and light curves of the source are obtained using data from the
two high-energy instruments 
JEM-X1 \citep{lun03} in the $3 - 20 \rm \, keV$ band and from IBIS/ISGRI
\citep{ube03} in the range $22 - 200 \rm \, keV$.   
The instrument data are  extracted for each individual SCW and
processed using the Off-line Scientific Analysis
(OSA v7.0) software released by the 
\integral\/ Scientific Data Centre (Courvoisier et al. 2003).
Following the standard analysis, we use the latest response matrix with 64 channels.
Then, data above 90 keV are rebinned to improve the signal to noise ratio.\\ 
The {\it  RXTE}/ASM (Levine et al. 1996)
daily averaged light curve, provided by the ASM/RXTE teams at 
MIT and at the RXTE SOF and GOF at NASA's GSFC,
(Fig. 1, panel a, from 
{\em {http://xte.mit.edu/ASM\_lc.html}}) 
shows that \source\/ has been continuously active since beginning 
of August for two months with multiple peaks.
The outburst of this transient source was frequently observed by 
{\it INTEGRAL} (Haymoz et al. 2007, Galis et al. 2007)
during the Key Programme on the Galactic Centre and private Target of Opportunity observations. 

The IBIS/ISGRI light curve (Fig. 1, panel b and c)
shows a gradual brightening in two energy band,
22-45 keV and 45-68 keV, while the ASM peak intensity was not monitored with \integral.

We report here on the outburst emission measured by the IBIS/ISGRI instrument
by dividing it in four separate epochs (see Table 1). These correspond to time periods  
during which the source spectra appear quite stable, with very small or absent spectral
variability as monitored on the time scale of a few SCWs.\\
We searched simultaneous IBIS and PCA (Glasser et al. 1994) data in the XTE public archive
{\footnote[2]{ 
{\em{http://heasarc.gsfc.nasa.gov/docs/xte/xhp\_archive.html}}
}},
but unfortunately 
only for epoch 1 PCA standard products are available.
No public PCA data are available for the epoch 2, 3 and 4 .
For our analysis we use the PCA standard products OBSID 93414-01-04-01.
Data are collected in standard2 modes with a time resolution of 16s and 129 energy channels
and from PCU 2 and PCU 4.
Source and background spectra are generated with {\it{SAEXTRCT}} version 4.2d
and response files with the tool {\it{PCARMF}} v10.1.
Background rates were estimated using the epoch-5 models, as provided by the PCA calibration team.
\\
\begin{table}
\tablecolumns{6}
\begin{center}
\label{tbl1}
\begin{tabular}{cccccc}
\hline\hline
&Instrument&Tstart &Tstop &   Exposure&counts/s$^a$\\
&&MJD    &MJD            &  ksec   &\\
\hline
1&IBIS&54337 &54357                          &      572  & 5.80$\pm$0.07              \\
2&IBIS&54358 &54362                          &      107   &  11.6$\pm$0.2                \\
3&IBIS&54367 &54370                          &      121   &  10.9$\pm$0.2            \\
4&IBIS&54373 &54376                          &      128   &  10.2$\pm$0.1               \\
5&PCA&54345.49&54345.51                     &      2      &  75.0$\pm$0.3           \\
\hline
\end{tabular}
\end{center}
\caption{Log of SAXJ1810.8-2609 IBIS and PCA observations.
}
$^a$ Rates are in the 22-200 kev energy range for IBIS  and  in 3-30 keV
for PCA  spectra. Source counts are background subtracted.
\end{table}

\section{SPECTRAL ANALYSIS}

\subsection{The wide band outburst emission}

The IBIS spectra extracted for the four epochs listed in Table~1 
have been fitted with both a simple power~law and a 
{\scriptsize{COMPTT}} model (Titarchuk 1994),
assuming a spherical geometry
for the Comptonizing
region.
Results are reported in Table~2.
The temperature of the Comptonizing
electrons kT${\rm _e}$ and the plasma optical depth
${\rm \tau_p}$ were free parameters in the fit, while the temperature of the soft 
photon Wien distribution kT${\rm _0}$ was fixed at 0.6 keV.
This is the value observed by \sax\/ in 1998 and previously reported by Natalucci et al. (2000).
Spectra are well described by a simple power law after the outburst peak, 
while a {\scriptsize{COMPTT}} model is required before the outburst peak.
Using a thermal Comptonization model {\scriptsize{COMPTT}} instead of simple power law 
did not give significantly better fits for epoch 3 and 4,
with the corresponding F-test chance 
probabilities being $6\times 10^{-2}$ and $3\times 10^{-3}$, respectively. 
Instead this model is statistically highly significant for the epoch 1 and 2, 
with the corresponding F-test chance
probabilities of $4\times 10^{-6}$ and $2\times 10^{-9}$, respectively.

For the first period we build a spectrum in a broad energy band (3-200 keV), 
using simultaneous IBIS and PCA data.
The most simple model which provides a good fit to this spectrum
is made up of a thermal comptonized component {\scriptsize{COMPTT}} in XSPEC 
(Titarchuk 1994) with 
a spherical geometry plus
a soft component consisting of a single temperature blackbody and an Gaussian component for iron line.\\
In the fitting procedure, a multiplicative constant has been introduced to take into account
possible cross calibration mismatches between the soft X-ray and the \integral\/ data;
this constant has been found to be 1.05$\pm$0.05.
Results are reported in Table~2. 
The iron line centroid is $6.3^{+0.8}_{-0.4}$ keV, $\sigma_{\rm Fe}<0.6keV$ and equivalent width $139^{+43}_{-50}eV$.
Figure \ref{res} shows four spectra and the
residuals with respect to the corresponding best fits.
Data and models are shown in {Figure \ref{fig3}, for four epochs.

\begin{table}
\label{tbl3} 
\tablecolumns{6}
\begin{center}
\begin{tabular}{lccccc}
\hline\hline
&&&&& \\
\multicolumn{6}{c}{IBIS spectra}\\
\multicolumn{6}{c}{Epochs 1, 2, 3, 4}\\
&&&&& \\
\hline
&&&&& \\
       & $\Gamma$
       & ${kT}_{e}$, $E_{c}$
       & $\tau$
       & $Flux_{20-100 keV}$
       &$\chi^2_\nu$ \\
       & 
       & keV
       & 
       & $_{10^{-10}erg~cm^{-2}~s^{-1}}$
       & $\chi^2$(d.o.f.)\\
&&&&& \\
\hline
\multicolumn{6}{c}{Power Law}\\
1	&$2.30_{-0.06}^{+0.10}$  		&... 		&... 	        &3.6    &1.43[30] \\ 
2	&$2.32_{-0.06}^{+0.05}$	           &...	 	&...    	&7.5    &1.49[29]\\
3	&$2.67_{-0.06}^{+0.07}$	            &...            &...    	&6.3    &1.13[27]\\
4	&$2.43_{-0.05}^{+0.06}$	            &...            &...    	&6.0    &0.92[32]\\
&&&&& \\
&&&&& \\
\multicolumn{6}{c}{\em comptt}\\ 
1   &...    & $30_{-7}^{+29}$  &  $1.2_{-0.7}^{+0.4}$   & 3.6  & 0.70[29]\\
2   &...    & $23_{-3}^{+8}$   &  $1.5_{-0.5}^{+0.4}$   & 7.5  & 0.40[28]\\
3   &...    & $69_{-4}^{+4}$   &  $<0.8$                & 6.3  & 1.02[26]\\
4   &...    & $87_{-5}^{+5}$   &  $<0.8$                & 6.0  & 0.81[31]\\
&&&&& \\
\hline
&&&&& \\
\multicolumn{6}{c}{IBIS and PCA spectra}\\
\multicolumn{6}{c}{Epoch 1}\\
&&&&& \\
\hline
&&&&& \\
       &${kT}_{bb}$
       & ${kT}_{e}$
       & $\tau$
       & $Flux_{3-100 keV}$
       & $\chi^2_\nu$\\
       & keV
       &keV
	&
       & $_{10^{-10}erg~cm^{-2}~s^{-1}}$
       & $\chi^2$(d.o.f.)\\
&&&&& \\
1   &$0.44\pm0.06$         &$22\pm3$              &$1.7\pm0.3$  & 8.1 & 1.05[78] \\
&&&&& \\
\hline\hline
\end{tabular}
\caption{Parameter values of spectral models fitting the outburst 
emission in the energy range 22-200keV using IBIS data and in the energy range 3-200keV 
using PCA plus IBIS data. Uncertainties are given at a 90 \%
confidence level. }
\end{center}
\end{table}

\begin{figure}[h]
\begin{center}
\includegraphics[width=10cm, angle=-90]{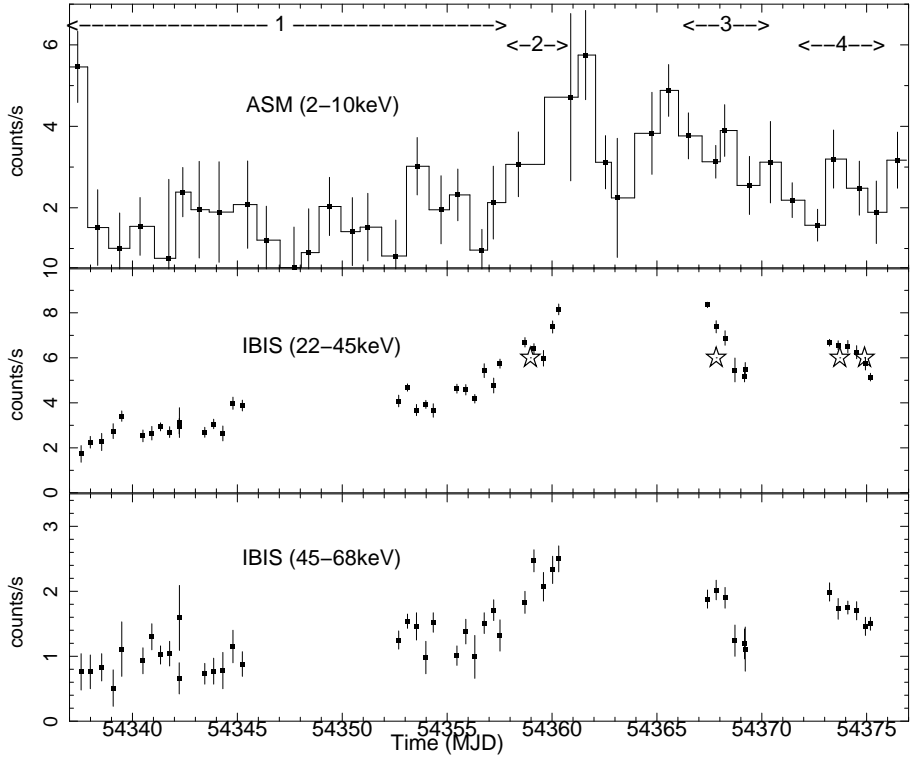}
\end{center}
\caption{a)The RXTE ASM light curve daily averaged in 2-10 keV energy band of
SAX~J1810.8-2609. b) The IBIS light curve  in 22-45 keV energy band. 
Stars indicate only burst times  c) The IBIS light curve  in 45-68 keV energy band.
  \label{fig1}}
\end{figure}

\subsection{The bursts emission}
In Fig. \ref{fig:burst1} and \ref{fig:3bursts} we show the  
JEM-X 3--25 keV light curves for the four bursts. 

The start time for each burst was determined when the intensity rose
to 
10~\% 
above the persistent intensity level. 
The rise time is defined as the time between the start of the burst and 
the time at which the intensity reached 90~\% of the peak burst intensity, 
as measured from the 2~s bin light curve in the full 3--25~keV band.
The burst duration is the approximate time it takes to the 3--25~keV intensity 
(averaged over 3 consecutive bins) to decrease back to the average persistent level 
previous to the burst start.

The spectral analysis of the bursts is based on JEM-X data in the 3--25~keV band.
Unfortunately, time resolved spectral analysis of such short bursts, requiring 
relatively high time resolution, leads to statistically poor results 
due to the little aperture of the JEM-X instrument.
Therefore,
a time-averaged spectral analysis over the first 18~s including the peak 
has been performed for each burst and 
every burst spectrum is well fit  
by a simple blackbody model. 

\begin{table}[htb] 
\caption{Analysis results of the bursts} 
\label{table:spec1}
\begin{center} 
\begin{tabular}{lllll} 
\hline \hline
Dataset  & Burst 1 & Burst 2 & Burst 3 & Burst 4 \\ 
\hline 
\noalign{\smallskip} 
Date (YYYYMMDD)  & 20070915 & 20070924 & 20070930 & 20071001 \\
Start time (UTC) & 23:20:18 & 19:52:50 & 17:10:03 & 21:42:10 \\
$kT_{\rm bb}$ (keV) & 1.4$^{+0.4}_{-0.3}$  & 2.1$^{+0.2}_{-0.2}$ 
 &  2.6$^{+0.3}_{-0.3}$ & 2.5$^{+0.5}_{-0.4}$\\ 
$R_{\rm bb, d_{5.5 kpc}}$ (km) & 18$^{+20}_{-6}$ & 6.1$^{+2}_{-1}$ & 
4.1$^{+2}_{-1}$  &   4.5$^{+2}_{-1}$\\ 
$\chi^{2}/{\rm dof}$  &  17/27 & 40/43 &  30/24 & 31/57\\   
$F_{\rm bol}$ \small $^{a}$ & $4.5\pm2.3$ & $2.8\pm0.8$ & $2.7\pm0.7$ & $3.0\pm1.3$ \\ 
\hline 
\multicolumn{2}{l}{Burst parameters} \\
$F_{\rm peak}$ \small $^{a}$ & $9.6\pm1.9$ &  $5.0\pm0.6$ & $6.8\pm0.7$ & $5.5\pm1$ \\
$f_{\rm b}$ \small $^{b}$ & $1.1\pm0.7$ & $0.7\pm0.2$ & $0.8\pm0.2$ & $0.7\pm0.3$ \\
Rise time ($\pm1$ s)  & 5 & 3 & 4 & 7\\
Duration ($\pm2$ s)  & 30 & 30 & 30 & 25\\
$\tau$ \small $^{c}$   & $12\pm9$ & $14\pm5$ & $12\pm4$ & $13\pm8$\\
$\gamma$ \small $^{d}$ $(10^{-2})$ &  $0.9\pm0.2$  &  $1.0\pm0.2$  &  $0.9\pm0.2$  &  $0.8\pm0.2$ \\
\hline
\end{tabular}
\end{center} 
\small $^{a}$ Unabsorbed flux (0.1--100 keV) in units of $10^{-8}$erg cm$^{-2}$ s$^{-1}$. 
\small $^{b}$ Fluence ($10^{-6}$erg cm$^{-2}$). \linebreak
\small $^{c}$ $\tau (sec) \equiv f_{\rm b}/F_{\rm peak}$.
\small $^{d}$ $\gamma \equiv F_{\rm pers}/F_{\rm peak,Max}$; 
$F_{\rm pers}$ is the persistent flux in 0.1--100 keV energy range previous to the time of each burst,
$F_{\rm peak,Max}$ is the highest burst peak flux, here $F_{\rm peak}$ of Burst 1.
\end{table}

The inferred blackbody 
temperature, $kT_{\rm bb}$, and apparent blackbody radius at 5.5~kpc, $R_{\rm bb}$, 
for every burst are reported in Table \ref{table:spec1}.
Burst fluences are obtained from the bolometric fluxes, $F_{\rm bol}$, 
extrapolated in the 0.1--100 keV energy range and integrated over the 
respective burst durations.
The peak fluxes, $F_{\rm peak}$, are obtained by comparing the peak count rate of
the 2~s bin light curves with the time-averaged count rate of the spectra medeled with a blackbody (see Table 3). 
Bolometric fluxes are extrapolated 
between 0.1 and 100 keV using XSPEC software.
All uncertainties in the spectral parameters are given at a 90 \% 
confidence level.

The first burst has the highest peak flux reaching a value of 
$\simeq500~cts/s$ corresponding to $\approx3.5~Crab$ 
($1~Crab \approx3\times 10^{-8} erg~cm^{-2}~s^{-1}$ between 3--25 keV).
From the light curve, it even seems to be preceded by a precursor 24~s in advance. 
However, since this burst was observed close to the limit 
of the JEM-X field of view, the significance of the precursor is quite low and 
therefore its reality may be doubtful.
Assuming that the main peak flux corresponds to the Eddington limit 
for a helium burst, $L_{\rm Edd}\approx\,3.8\times 10^{38}$ erg s$^{-1}$, 
as empirically derived by Kuulkers et al. (2003), we can calculate an upper limit to 
the distance of the source of about 5.7~kpc. 
For comparison, 
the theoretical value of $L_{\rm Edd}=2.9\times 10^{38}$ erg s$^{-1}$,
assuming a helium atmosphere, a canonical mass of $1.4 \rm \, M_\odot$ 
and 10 km radius 
for the neutron star 
photosphere
(e.g. Lewin et al. 1993), leads to a distance of 5.0 kpc, 
consistent with the distance previously derived by Natalucci et al. (2000). 
The three remaining bursts are all weaker,
reach approximately the same peak flux, and have similar decay times.
From the detection of four bursts during
the total observation time of about 928~ks elapsed on the source by {\it INTEGRAL},
due to the non continuous coverage of the outburst, 
we can estimate an approximate recurrence time of 2.7 days in average.
Nevertheless, as the fourth burst occurred the day after the third burst, 
namely $\Delta t_{\rm 3-4} = 102730~s$ = 1.2 days later, 
this interval represents a more stringent constrain on the bursting rate (see below).

The average persistent unabsorbed flux between 0.1--100 keV, 
$F_{\rm pers}\approx~5\times 10^{-10}$ erg cm$^{-2}$ s$^{-1}$,
translates to a bolometric luminosity 
$L_{\rm pers}\approx~1.8\times 10^{36}$ erg s$^{-1}$, 
assuming an approximate distance of 5.5 kpc. 
This corresponds to a mass accretion rate per unit area 
equal to
$\dot m=L_{\rm pers}(1+z)\;\eta^{-1}$ c$^{-2}/A_{\rm acc} 
\approx~10^{3}$ g cm$^{-2}$ s$^{-1}$ 
(where $A_{\rm acc}= 4\pi R_{\rm NS}^{2}$ and 
$\eta = GM_{\rm NS}/(R_{\rm NS}$~c$^{2}) \simeq~0.2$ 
is the accretion efficiency for a canonical neutron star).

\begin{figure}[t]
\begin{center}
\includegraphics[width=5cm, angle=-90]{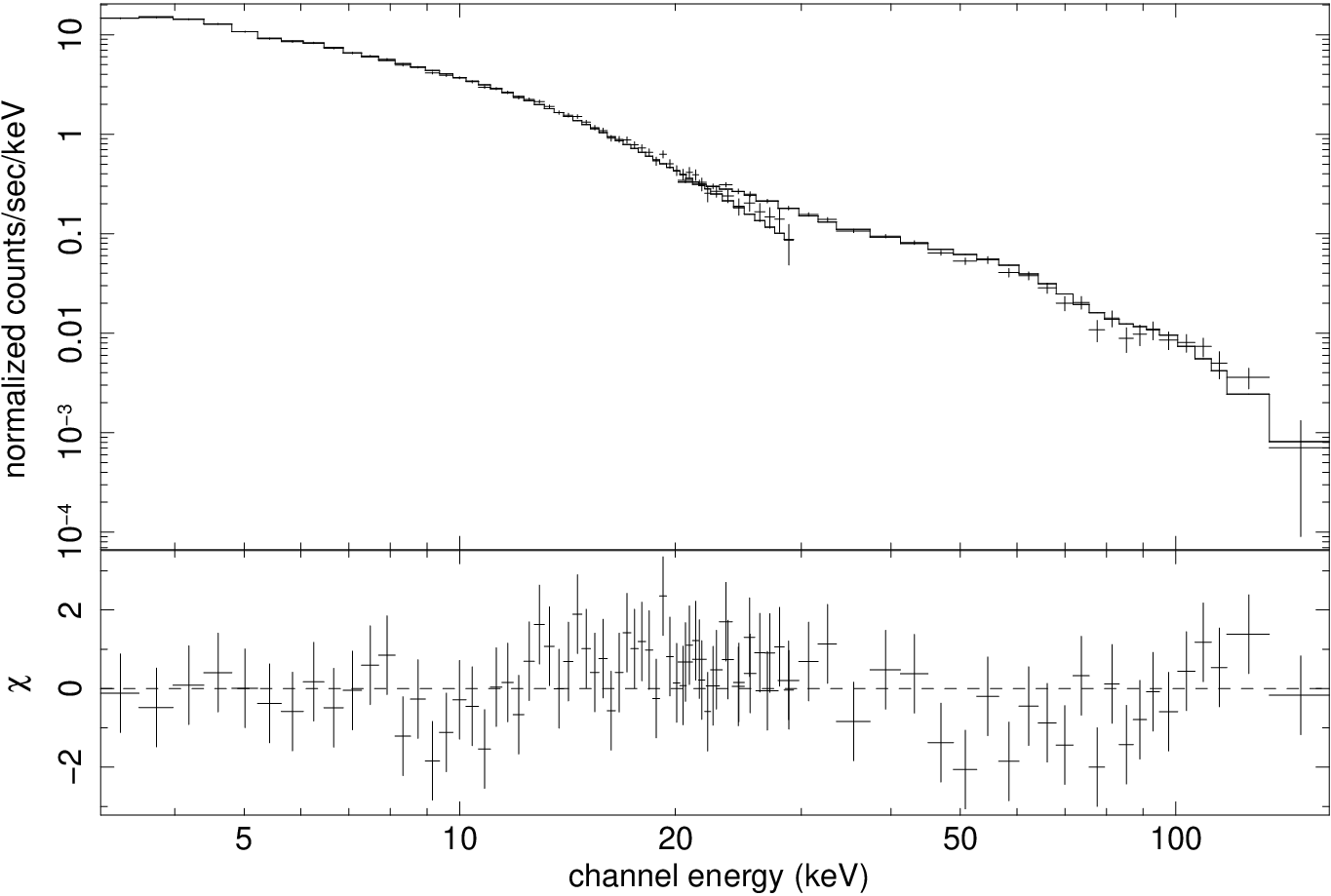}
\includegraphics[width=5cm, angle=-90]{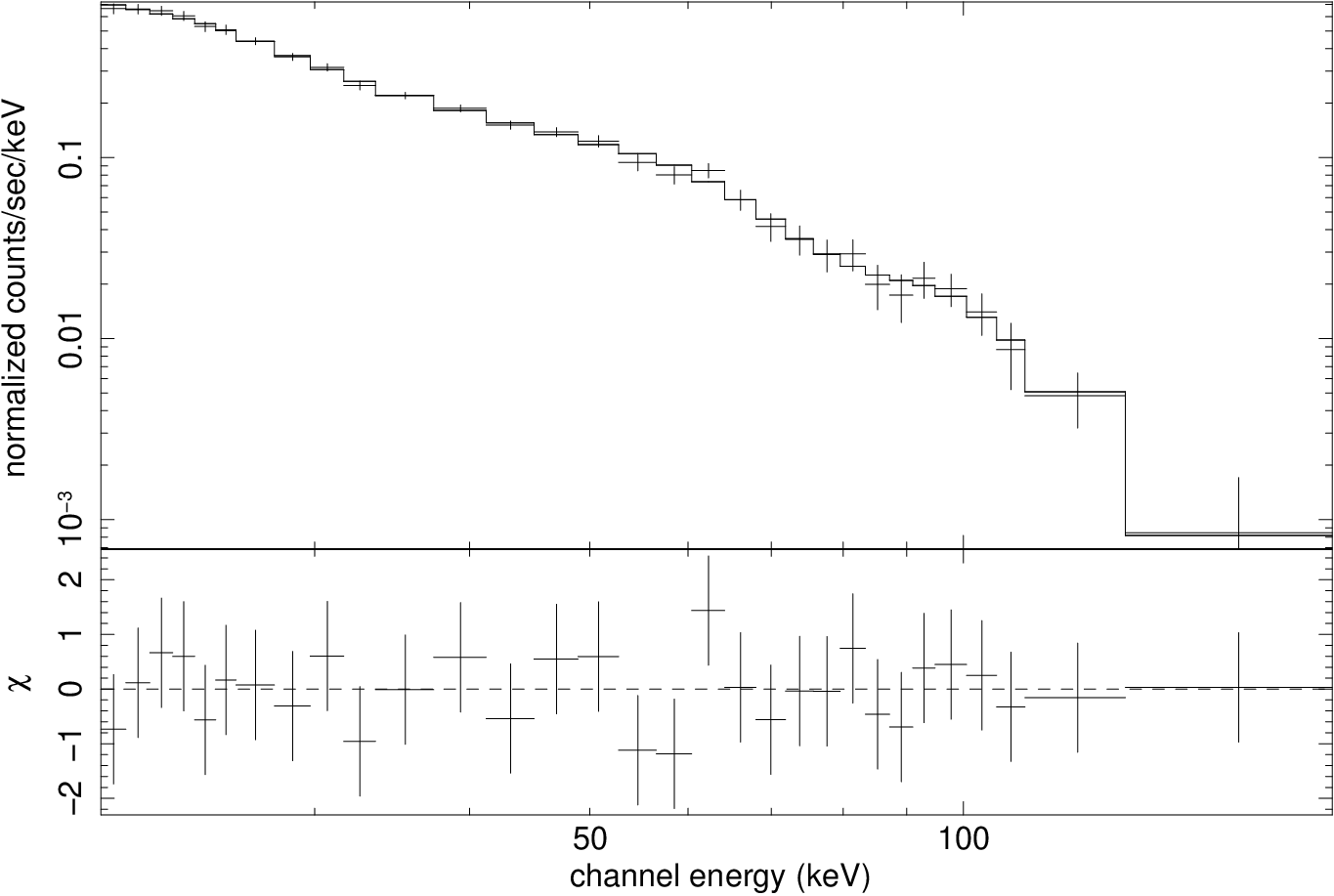}
\includegraphics[width=5cm, angle=-90]{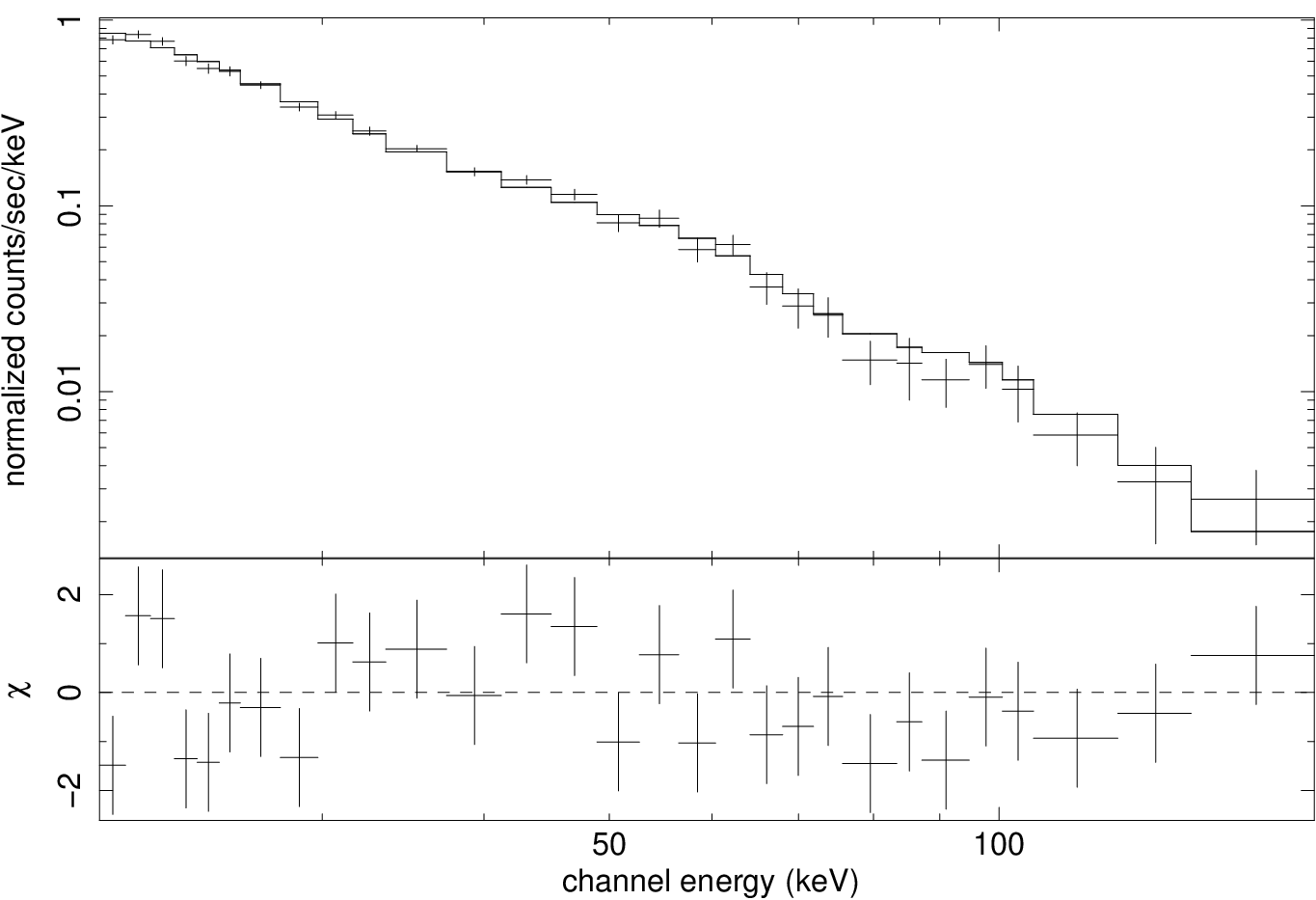}
\includegraphics[width=5cm, angle=-90]{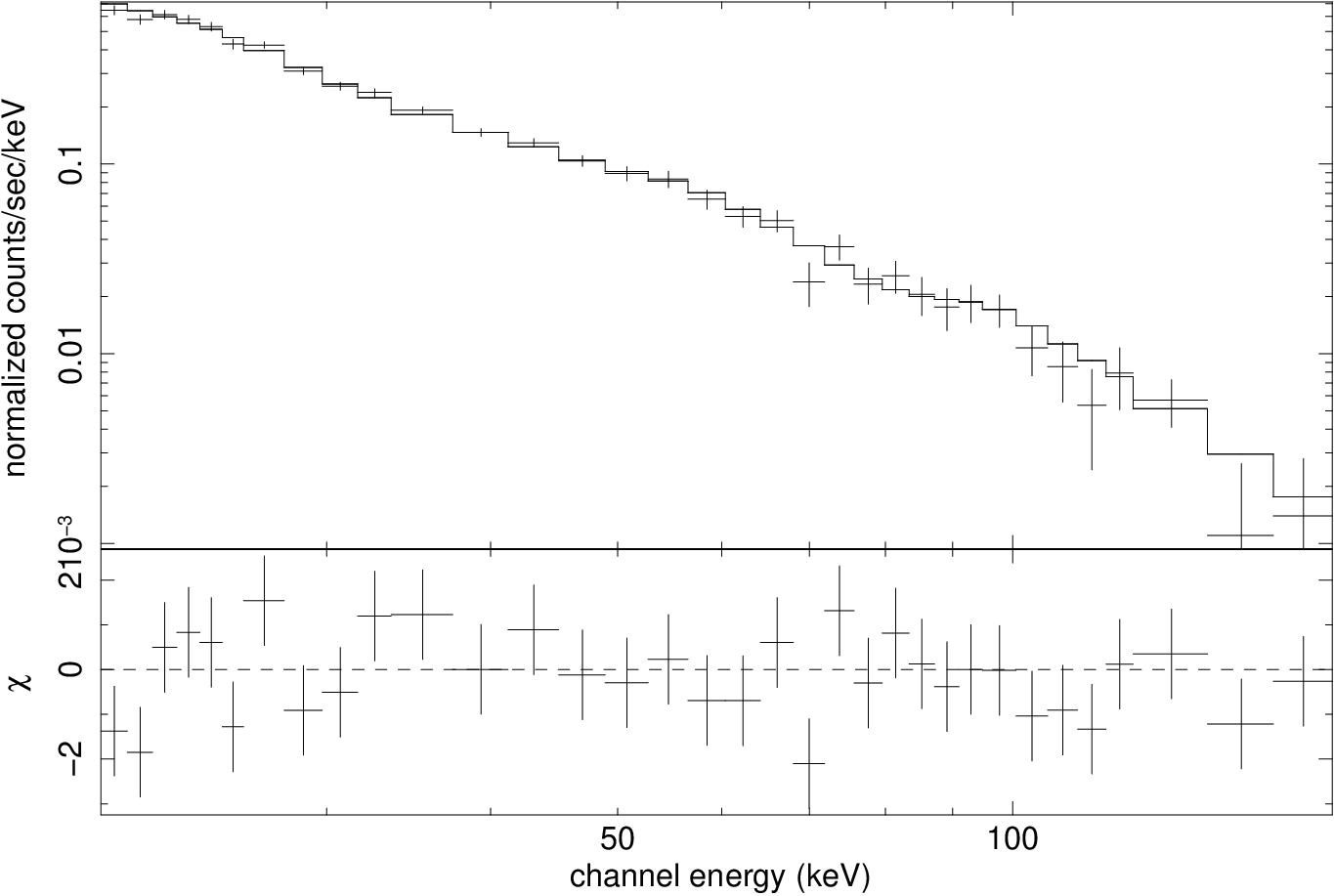}
\end{center}
\caption{Four spectra of \source\/ and the
residuals with respect to the corresponding best fits,  {\scriptsize{COMPTT}}
plus blackbody component for epoch 1, simple {\scriptsize{COMPTT}} model for epoch 2 and 
simple power law component for epochs 3 and 4.
For the epoch 1 spectrum is obtained using PCA and IBIS data (panel 1), while for others epochs
spectra are obtained using only IBIS data (panels 2, 3 4).
\label{res}}
\end{figure}

\begin{figure}[h]
\begin{center}
\includegraphics[width=10cm, angle=-90]{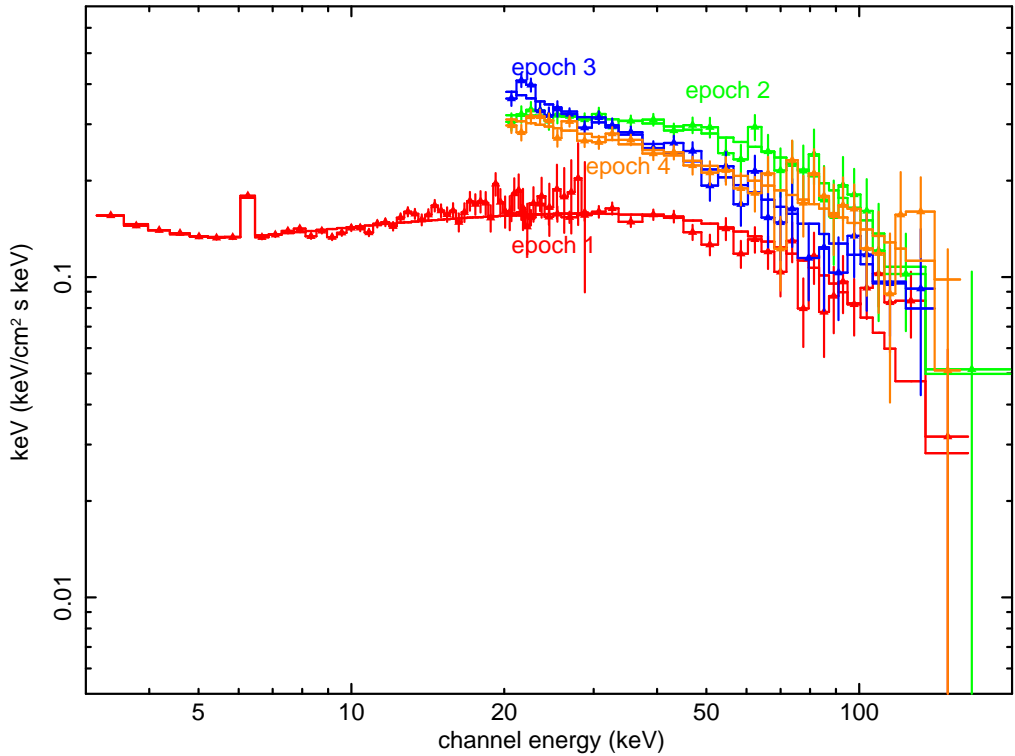}
\end{center}
\caption{Four different spectral states with the {\scriptsize{COMPTT}} model.
For the epoch 1 PCA and IBIS spectra are fitted (red color), while for others epochs
only IBIS spectra are fitted (green, blue and orange colors).
}
  \label{fig3}
\end{figure}

The total energy released by the first burst was 
$E_{b,1}\simeq 4\times 10^{39}$ erg 
which, assuming complete and isotropic burning, corresponds to an ignition column 
$y = E_{b,1}(1+z)/4\pi R_{\rm NS}^{2} Q_{\rm nuc}$  
ranging between $y\approx~1\times10^{8}$ g cm$^{-2}$ for 
burning hydrogen with abundance X=0.7,
and $y\approx~2.6\times10^{8}$ g cm$^{-2}$ for X=0 (pure helium); 
here $Q_{\rm nuc} = 1.6 + 4X$ MeV/nucleon 
is the nuclear energy release for a given average hydrogen fraction  
at ignition X, and z=0.31 is the 
appropriate gravitational redshift at the surface of a 1.4 M$_{\odot}$ 
neutron star (Cumming, 2003).
From the relation $\Delta t_{\rm rec}= y(1+z)/\dot m$ a burst
recurrence time of $1.5$ days is expected for X=0.7, 
and $\Delta t_{\rm rec}=3.8$ days for pure helium burning.
The same calculations for the fourth burst with an energy release of
$E_{b,4}\simeq 2.5\times 10^{39}$ erg lead to 
$\Delta t_{\rm rec}=0.9$ days for X=0.7, and $\Delta t_{\rm rec}=2.3$ days for X=0. 
The observed recurrence time seems thus most consistent with 
mixed H/He burning.

\begin{figure}[htb] 
\begin{center}
\includegraphics[width=10cm, angle=-90]{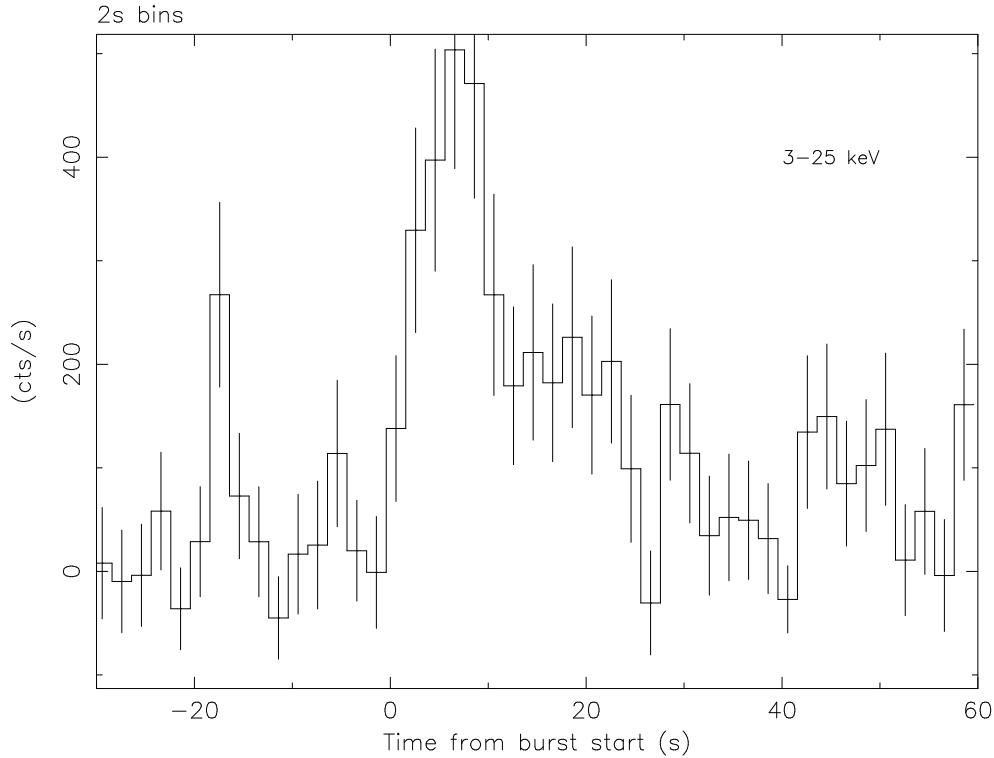} 
\end{center}
\caption 
{The first type I X-ray burst detected on September 15, 2007.
Time 0 corresponds to 23:20:18 (UTC). 
The JEM-X (3--25 keV) light curve is shown with a time bin of 2 s.
}  
\label{fig:burst1} 
\end{figure} 

\begin{figure}[htb] 
\begin{center}
\includegraphics[width=10cm, angle=-90]{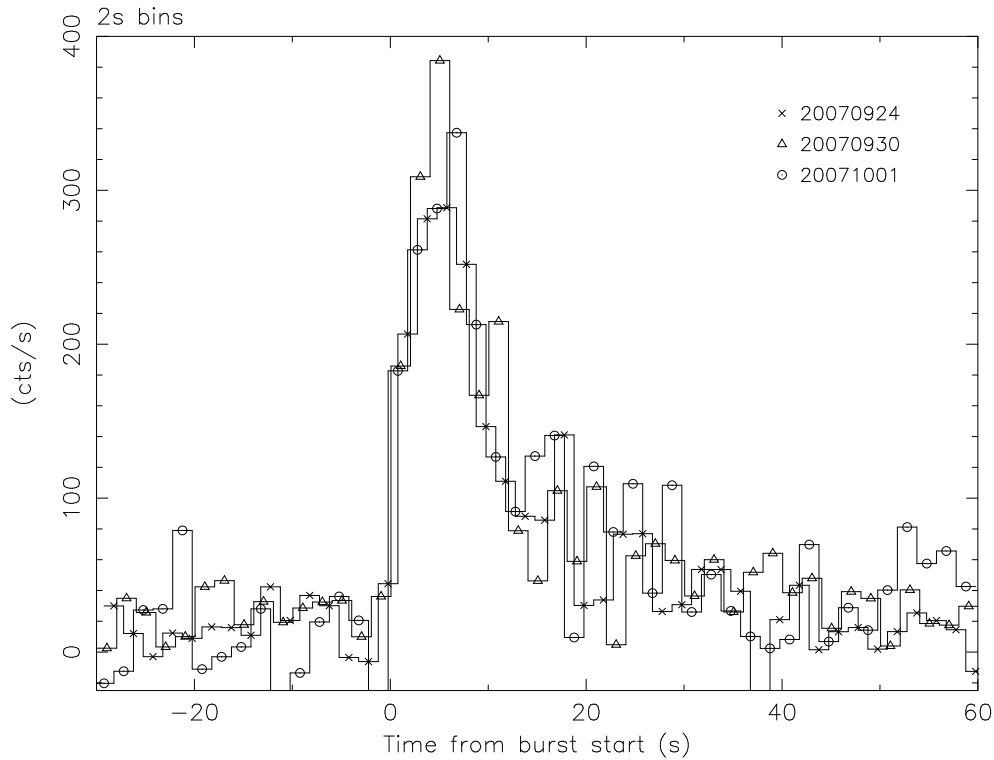} 
\end{center}
\caption 
{The three other bursts respectively detected on September 24, 30, and October 1st, 2007.
The JEM-X (3--25 keV) light curves are shown with a time bin of 2 s.}  
\label{fig:3bursts} 
\end{figure} 

Moreover, it is also possible to calculate the burst energetics by 
the ratio of the total energy emitted in the persistent flux
to that emitted in the bursts (e.g. Galloway et al., 2004):
$\alpha= (F_{\rm pers}/f_{b})\Delta t_{\rm rec}\approx 73$, 
for $\Delta t_{\rm rec}= \Delta t_{\rm 3-4}$, 
F$_{\rm pers}=5\times10^{-10}$ erg cm$^{-2}$ s$^{-1}$, 
and $f_{b}\simeq 0.7\times10^{-6}$ erg cm$^{-2}$ is the fluence of the fourth burst. 
Assuming that all the accreted fuel is burned during the bursts, 
the calculated $\alpha$-value from the measurable quantities is consistent with  
$Q_{\rm nuc}= (1+z) \eta c^2/\alpha\, (10^{18}\, erg/g)^{-1}\simeq$ 3.2 MeV/nucleon, 
corresponding to an hydrogen fraction X=0.4.
Since other bursts could have been burnt during the observation gaps, the calculated $\alpha$ value 
is only an upper limit and conversely the calculated value of X is a lower limit.
We can indeed conclude that the burst fuel could be composed by mixed hydrogen and helium with X$\geq$0.4.
 
\section{DISCUSSION}
The IBIS/ISGRI observations have allowed us to follow the high energy behavior 
of \source\/ during its long and bright X-ray outburst.
Light curves varied simultaneously in all
X-ray bands monitored (2--10 keV, 22--45 keV, 45--68 keV).
The X-ray spectra appeared always to be comparably soft, with a photon index 
of $\Gamma=  2.3-2.7$. 
The 20--100 keV luminosities are in the range $L_X =
1.1-2.6~\times~10^{36} \rm \, erg \, s^{-1}$ (estimated for a source at 5.5 kpc),
which is typical for the low hard state of neutron star binaries (Barret et al.~2000).
We estimated a fluence
of $5.5\times10^{-3}erg~cm^{-2}$ using an 
average bolometric flux of $1.6\times10^{-9}erg~cm^{-2}~s^{-1}$ in the 0.1-500 keV
energy band 
and an outburst duration of $\sim$40~days from ASM light curve.  
The long term time averaged accretion rate
is $\dot{M}\simeq 5 \times 10^{-12} {\rm M}_{\sun}$ yr$^{-1}$, 
taking into account a time interval between two outbursts of $\sim$9.6yr.
The ASM, IBIS and BAT
{\footnote[3]{in the 15-50 energy range, from 
{\em{http://heasarc.gsfc.nasa.gov/docs/swift/results/transients/weak/SAXJ1810.8-2609}}}} 
light curves 
show the same behavior, so the outburst duration and the time between two outbursts
are estimated using the ASM light curve.
This time-average low mass accretion rate, the outburst luminosity 
of $\sim1-3\times10^{36}~erg~s^{-1}$
lower than typical values for neutron star soft X-ray transient 
($\sim2\times10^{38}~erg~s^{-1}$),
together with low quiescent luminosity ($\sim1\times10^{32}~erg~s^{-1}$)
reported by Jonker et al. 2004, strengthens the idea of these authors that this 
source belongs to the class 
of faint soft X-ray transient.
In addition we note that the low average persistent bolometric luminosity
is very similar as the luminosity of the Ultra Compact X-ray Binaries (see, e.g.,
Fiocchi et al. 2008 and Falanga et al., 2008).
However, we think difficult to derive a conclusion about the ultra compact nature of the system
because the derived hydrogen fraction in the burst fuel of \source\/ is not consistent with an
ultra compact source, since those are thought to accrete pure Helium from a white dwarf (Nelemans and Jonker, 2006).

Spectral parameters are not correlated with the observed luminosities,
but instead, they vary according with the rise/decay phases of the outburst.
During the rise of the flux,
the \source\/ luminosity changes by a factor 2, while there are not modifications of the spectral shape:
the electron temperature $kT_e$ is $\sim$ 23-30 keV
and optical depth $\tau$ of the plasma is $\sim$ 1.2-1.5.
This hard X-ray emission could be interpreted in the standard way, as produced by the
upscattering of soft seed photons by a hot, optically thin electron plasma. 
During the decrease of the flux, spectra show a harder spectral shape
with an optical depth of the plasma lower than 0.8 and  very high 
electrons temperatures $kT_e$ of $\sim$ 69-87 keV.
The spectral parameters measured during the decay phase of the 2007 outburst
agree with the ones found using the \sax\/ observations (Natalucci et
al. 2000), showing the same X-ray spectral behavior: during the decay
phase of the outbursts of 1998 also no high energy spectral steepening was observed. 
\\
We cannot determine whether the emission is due to a thermal or non-thermal process,
because equally good fits are obtained either with a power law with no detectable cutoff 
below $\sim$100 keV or with a thermal Comptonization spectrum with an electron 
temperature in excess of $\sim$ 80-90 keV.
This electron distribution
could also arise from Comptonization by hybrid (thermal and non thermal) corona
(Coppi 1999), or from the Compton cloud located inside the neutron stars magnetosphere
(Titarchuk et al. 1996),
or, alternatively, the power-law component could be produced by
Comptonization of synchrotron emission in a relativistic jet 
(Bosch-Ramon et al. 2005, Fender 2004).
Up to date there are few detections of radio emission associated 
with neutron star X-ray transients and sometimes outbursts of soft X-ray transient were associated 
with strong transient radio emission (Ball et al. 1995, Kuulkers et al. 1999, Fender \& Kuulkers 2001). 
A comparison with hard tails detected
from neutron star systems
and some black hole
binaries could be interesting suggesting that a similar mechanism could originate these components.

\acknowledgements
We acknowledge the ASI financial/programmatic support via contracts ASI-IR
I/008/07/0.
JC acknowledges financial support from ESA-PRODEX, Nr. 90057.
A very particular thank to K. Hurley for making data
available before becoming public.
Finally, we thank the anonymous referee for the constructive comments and detailed review of the paper.

%
%

\end{document}